\documentclass{article}
\usepackage[utf8]{inputenc}
\usepackage{algorithm2e}
\usepackage{amsmath,color}
\usepackage{amssymb}
\usepackage{natbib}
\usepackage{graphicx}
\usepackage{subfig}
\usepackage{mathtools}
\usepackage{multirow, setspace, hyperref}
\usepackage{array}
\usepackage{float}
\usepackage{tcolorbox}
\usepackage{caption}
\usepackage{subcaption}

\hypersetup{hidelinks}

\usepackage[top=1in, bottom=1in, left=1in, right=1in]{geometry}
\newcolumntype{L}{>{$}l<{$}} % math-mode version of "l" column type

\usepackage{titlesec}
\usepackage{enumitem}
\usepackage{etoolbox}

\BeforeBeginEnvironment{algorithm}{\vskip-1ex}
\AfterEndEnvironment{algorithm}{\vskip-1ex}
\usepackage{scalerel}
\usepackage{stackengine,wasysym}

\newcommand{\indep}{\perp \!\!\! \perp}

\titleformat*{\section}{\large\bfseries}
\titleformat*{\subsection}{\normalsize\bfseries}
\titleformat*{\subsubsection}{\normalsize\bfseries}

\usepackage[font=normalsize,skip=0pt]{caption}
\usepackage{fontsize}
\title{The Conditional Prediction Function: A Novel Technique to Control False Discovery Rate for Complex Models}
\author{Yushu Shi, Michael Martens}
\date{\today}
\begin{document}
%	\setlength{\parindent}{0em}
%	
%	\setlength{\abovedisplayskip}{0.5pt}
%	\setlength{\belowdisplayskip}{0.5pt}
%	\setlength{\abovedisplayshortskip}{0.5pt}
%	\setlength{\belowdisplayshortskip}{0.5pt}
%	
%	%\titlespacing{\section}{0pt}{\parskip}{-\parskip}
%	%\titlespacing{\subsection}{0pt}{\parskip}{-\parskip}
%	%\titlespacing{\subsubsection}{0pt}{\parskip}{-\parskip}
%	
%	\titlespacing\section{0pt}{8pt plus 4pt minus 2pt}{0pt plus 2pt minus 2pt}
%	\titlespacing\subsection{0pt}{8pt plus 4pt minus 2pt}{0pt plus 2pt minus 2pt}
%	\titlespacing\subsubsection{0pt}{8pt plus 4pt minus 2pt}{0pt plus 2pt minus 2pt}
%	
%	
%	\setlist[itemize]{noitemsep, topsep=0pt}
%	\setlist[enumerate]{noitemsep, topsep=0pt}
%	\setlist[description]{noitemsep, topsep=0pt}
%
%

\maketitle

\begin{abstract}
In modern scientific research, the objective is often to identify which variables are associated with an outcome among a large class of potential predictors. This goal can be achieved by selecting variables in a manner that controls the the false discovery rate (FDR), the proportion of irrelevant predictors among the selections. Knockoff filtering is a cutting-edge approach to variable selection that provides FDR control. Existing knockoff statistics frequently employ linear models to assess relationships between features and the response, but the linearity assumption is often violated in real world applications. This may result in poor power to detect truly prognostic variables. We introduce a knockoff statistic based on the conditional prediction function (CPF), which can pair with state-of-art machine learning predictive models, such as deep neural networks. The CPF statistics can capture the nonlinear relationships between predictors and outcomes while also accounting for correlation between features. We illustrate the capability of the CPF statistics to provide superior power over common knockoff statistics with continuous, categorical, and survival outcomes using repeated simulations. Knockoff filtering with the CPF statistics is demonstrated using (1) a residential building dataset to select predictors for the actual sales prices and (2) the TCGA dataset to select genes that are correlated with disease staging in lung cancer patients.
\end{abstract}

\section{BACKGROUND}
The frequency and sizes of high dimensional datasets have increased greatly in recent years, accompanied both by new opportunities to find determinants of outcomes of interest and by formidable challenges to identify truly prognostic variables among this myriad of candidates. Usually, many of them are not expected to affect the outcome and so interest lies in choosing only the variables that are most likely to be associated with the response. An appropriate way to formalize this objective is to focus on controlling the FDR \citep{BenjHoch95}. %Penalized regression models, such as the lasso \citep{Tibs96}, are often employed for feature selection; however, performing formal hypothesis testing of the effects of predictors and controlling the FDR are generally quite challenging for these procedures. 
Knockoff filtering \citep{BarbCand15} is a popular device for FDR controlled variable selection, particularly since this control is guaranteed for finite sample sizes rather than appealing to asymptotic properties for its validity, as other frequently used approaches do. The general problem involved is to determine, within a vector $\mathbf{X}$ of $p$ features, which ones are associated with the outcome $Y$ of interest. The commonly used Model-X knockoff approach \citep{CandFan18} conducts this evaluation as follows: 
\begin{itemize}[leftmargin=*]
	\item First, construct knockoff copies $\widetilde{\mathbf{X}}$ satisfying the following two conditions:
	\begin{enumerate}[leftmargin=*]
		\item \textbf{Pairwise exchangeability}: for any subset $S \subset \{1,\dots,p\},$ $(\mathbf{X},\widetilde{\mathbf{X}})_{swap(S)}\overset{d}{=}(\mathbf{X},\widetilde{\mathbf{X}})$, i.e. swapping any subset of variables with their knockoffs leaves the joint distribution invariant;
		\item \textbf{Conditional independence}: $\widetilde{\mathbf{X}} \indep Y|\mathbf{X}.$  
	\end{enumerate}
	\item Second, a model is fitted using all original and knockoff features, and the influences of the original and knockoff features are evaluated using importance statistics, such as the lasso \citep{Tibs96} penalty parameter at which a feature enters the model. The knockoff statistic built on it is the lasso signed max (LSM) statistic. Because the knockoff features are independent of the outcome, their importance statistics represent noise variables that can be contrasted against the  original features' to assess which original features correlate with the response. From the importance statistics of the original feature and its knockoff, $Z_j$ and $\widetilde{Z}_j$, $j=1,\dots,p$, we can compute knockoff statistics $W_j=f_j(Z_j,\widetilde{Z}_j),$ where $f_j$ is an anti-symmetric function. A large positive $W_j$ provides evidence that the distribution of $Y$ depends on $X_j,$ whereas under the null $W_j$ is equally likely to take on positive and negative values (the "coin-flip" property).
	\item   Third, features are selected based on the knockoff statistics in a manner that controls the FDR at the specified level, $q$. Specifically, if we find a threshold $\tau>0$ by setting $$
	\tau=\min\left\{t>0: \frac{\# \{ j:W_j\leq -t\} }{\# \{  j: W_j \geq t\}} \leq q\right\}.
	$$ Then the procedure selecting the set $$
	\hat{S}=\{j:W_j\geq \tau\}
	$$  controls the modified FDR as $$
	mFDR=\mathbb{E}\left[\frac{|\{j\in\hat{S}\cap \mathcal{H}_0\}|}{(|\hat{S}|+1/q)}\right]\leq q.
	$$ A more conservative procedure, knockoffs$+,$ with threshold $$\tau^+=\min\left\{t>0: \frac{1+\#\{ j:W_j\leq -t \} }{\{ \#  j: W_j \geq t\}} \leq q\right\}$$ controls the ordinary FDR at level $q$ by selecting $$
 	\hat{S}_+=\{j:W_j\geq \tau^+\}.
    $$
	
\end{itemize}

Previous work has explored various ways to improve knockoff feature generation with a guarantee of pairwise exchangeability. Notably, \cite{BateCand21} introduced a Metropolized knockoff sampler, which handles situations where the covariates are continuous, heavy-tailed, and follow an Ising model. Furthermore, other methods took advantage of generative machine learning models, including Deep Knockoffs \citep{RomaSesi19}, KnockoffGAN \citep{JordYoon19}, and variational autoencoder related methods \citep{LiuZhen18, DuarFeng20}. Compared with the abundance of choices for generating knockoff counterparts, few studies have attempted to improve knockoff statistics. Common choices of knockoff statistics are based on a linear regression based model, such as lasso \citep{BarbCand15, CandFan18}; a common choice is the lasso coefficient difference (LCD) statistic, which is defined as
$$
W_j^{LCD}=|\beta_j|-|\tilde{\beta}_j|,
$$
where $\beta_j$ and $\tilde{\beta}_j$ are the regression coefficients for $X_j$ and its knockoff $\tilde{X}_j$.
The lasso-based statistics can yield good power if the true relationship is linear \citep{WeinBarb17}. But, if the linearity assumption is violated, these approaches may have poor power to detect prognostic variables.
This paper investigates methods for constructing knockoff statistics that can yield higher power to resolve features with linear or nonlinear relationships with the outcomes while maintaining the FDR control. We propose the conditional prediction function (CPF) as a basis for building the knockoff importance statistic. The proposed CPF knockoff statistic is able to capture nonlinear relationships, and will empower researchers to explore more flexible prediction models for knockoff filtering, such as deep neural networks (DNNs).

In Section \ref{sec:Method}, we explain the rationale and summarize the algorithm for calculating CPF knockoff statistics. The CPF statistic's ability to capture non-linear relationship for different types of outcomes is demonstrated through repeated simulations in Section \ref{sec:Simulation}. Section \ref{sec:Application} applies the proposed method to a residential building dataset to select predictors for the actual sales prices and to the TCGA LUAD cancer dataset to identify genes that are associated with or more advanced stage of lung cancer. Section 5 concludes the paper with a short discussion.

\section{THE CPF STATISTIC}
\label{sec:Method}
In this section, we introduce the conditional prediction function (CPF), which can measure a feature's importance by changing its value by a small magnitude while conditioning on all other covariates and computing the prediction difference before and after the change. The concept of the CPF knockoff statistic originates from multivariable linear models, where the regression coefficient signifies how much the mean of the dependent variable changes given a one-unit shift in the independent variable while holding other variables in the model constant. The CPF can be paired with many complicated predictive models, however. In the following text, we assume that we have already generated high-quality knockoff features $\tilde{\mathbf{X}}$ from $p$ original predictors $\mathbf{X}.$ Calculation of the CPF knockoff statistics proceeds per the following:

\begin{enumerate}	
	\item Combine both the original and the knockoff predictors $\mathbf{X}^*_{[N\times 2p]}=[\mathbf{X},\tilde{\mathbf{X}}]$ in one matrix. Center and scale continuous predictors.
	\item Train the prediction model $\mathcal{M}$ for the outcome $Y$ with both the original and knockoff features.
	\item Set $U_i=0, \quad i=1,\dots,2p.$ Choose the number of percentiles at which we want to evaluate the CPF, $J$ , and the $\delta$ value for numerical calculation.  Then,
\end{enumerate}
\begin{algorithm}[h!]
%	\caption{CPF algorithm}
	\For{each predictor $X^*_i$, $i \leftarrow 1 $ \KwTo $ 2p$}{
		Sample a subset with $n$ observations. 
		
		\For{$k \leftarrow 1 $ \KwTo $ n$}{
			\If{$X_i^*$ is a continuous covariate}{
				\For{$j \leftarrow 1 $ \KwTo $ J$}{
					Denote the value of $X_i^*$ at the $100 j/(J+1)$ percentile as $X_{ij}^*$. Conditioning on the values of all other predictors $\mathbf{X}_{k,-i}^*$, evaluate the predictive difference for $X_{ij}^*-\delta/2$ and $X_{ij}^*+\delta/2$.
					
					$(\star)\quad U_i=U_i+\lVert\mathcal{M}(X_{ij}^*+\delta/2|\mathbf{X}_{k,-i}^*)- \mathcal{M}(X_{ij}^*-\delta/2|\mathbf{X}_{k,-i}^*))\rVert_2^2/(\delta^2 J)$
				}
			}
			\Else{
				$(\star)\quad U_i=U_i+||\mathcal{M}(X_{ij}^*=1|\mathbf{X}_{k,-i}^*)- \mathcal{M}(X_{ij}^*=0|\mathbf{X}_{k,-i}^*)||_2^2$
			}
		}
	}
	\For{each pair of the original predictor $X_m$ and the corresponding knockoff  $\tilde{X}_m$, $m \leftarrow 1 $ \KwTo $ p$}{
		$CPF_m=U_m-U_{m+p}$
	}

\end{algorithm}

	After computing knockoff statistics for each pair of the original and knockoff covariates, we implement the last step of knockoff filtering. Given the pre-specified FDR level $q$, we find the threshold value $T$ such that $$T=\min\left\{t>0:  \frac{\#\{r:CPF_r\leq -t\} }{ \#\{r:CPF_r\geq t\}} \leq q \right\},$$ and return the selected feature set $\hat{S}=\{r:CPF_r \geq T\}.$

%\begin{wrapfigure}{r}{0.5\textwidth}
\begin{figure}[t!]
	\begin{center}
		\includegraphics[width=0.45\linewidth]{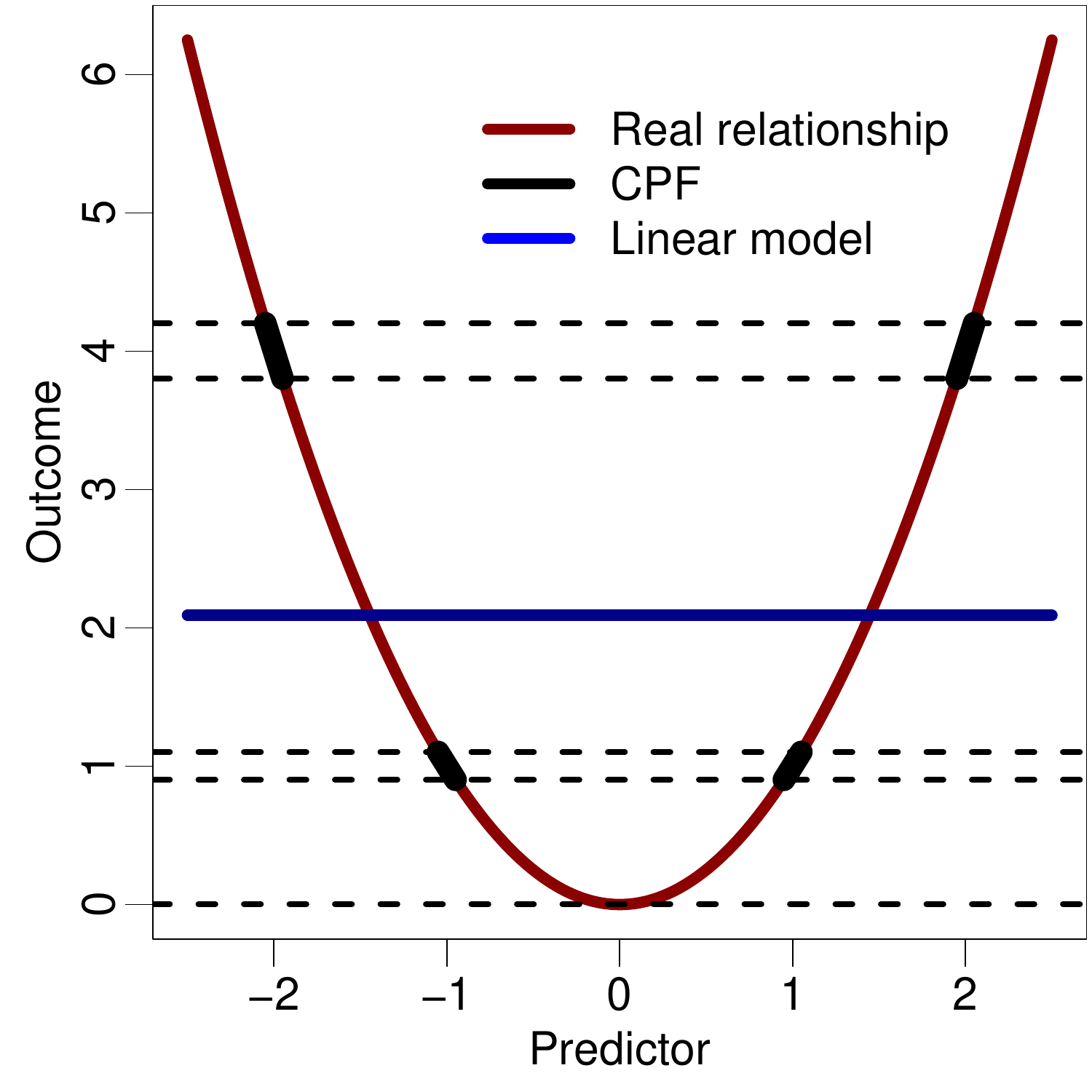}
	\end{center}
	\caption{CPF can quantify nonlinear covariate effect. Here the red curve is the true effect of the covariate. If we fit a linear model, the slope of the line will be 0, as indicated by the blue line. The black lines indicate the prediction difference that can be captured by CPF statistics.}
	\label{CPF}
\end{figure}
%\end{wrapfigure}

To illustrate the benefit of CPF, consider the univariate case where the predictor and the outcome have a relationship $y=x^2$. As shown in Figure \ref{CPF}, the predictor has a negative slope when less than $0,$ and a positive slope when greater than $0.$ If one uses a linear model or a linear model based method (such as lasso, elastic net), this variable will likely be declared null, as the positive effect when $x>0$ and negative effect when $x<0$ cancel out. But if one employs CPF as knockoff importance statistics, this nonlinear relationship will be well captured, since CPF measures the cumulative squared effect over the space of $x$.

When features are independent and the relationship to the outcome is genuinely linear, the CPF reduces to $n$ times the squared regression coefficient of a linear model. However, when the relationship is more complicated, by evaluating the conditional prediction differences at $J$ percentiles and averaging over $n$ possible combinations of other predictors, the CPF can overcome the rigidity of linear models.

When the outcome of interest is continuous, the predicted outcome $\mathcal{M}(\cdot)$ is a real number. When the outcome of interest is categorical, $\mathcal{M}(\cdot)$ is a vector of probabilities belonging to each category, and the increment in each iteration is the squared Euclidean distance of two probability vectors. When the outcome of interest is a survival time in the single failure cause setting, the CPF can be calculated over all the distinct time points in the dataset and $\mathcal{M}(\cdot|t)$ is the predicted survival function indexed by time $t$. When multiple failure types, i.e. competing risks, are present, $\mathcal{M}(\cdot|t)$ is the predicted cumulative incidence function (CIF) for the event type of interest. When the number of distinct event times $N_t$ is large, we can select a subset of $n_t$ time points (say, quintiles of the observed survival times), $\{t_l;l=1,\dots, n_t\}$. For survival outcomes, the $\star$ step in the algorithm becomes 

\begin{algorithm}[h!]
	\For{$l \leftarrow 1 $ \KwTo $ n_t$}{
		$U_i=U_i+||\mathcal{M}(X_{ij}^*+\delta/2|t_l,\mathbf{X}_{k,-i}^*)- \mathcal{M}(X_{ij}^*-\delta/2|t_l, \mathbf{X}_{k,-i}^*)||_2^2/\delta^2J$ 
	}
\end{algorithm}

if $X*_i$ is continuous, and 

\begin{algorithm}[h!]
	\For{$l \leftarrow 1 $ \KwTo $ n_t$}{
		$U_i=U_i+||\mathcal{M}(X_{ij}^*=1|t_l,\mathbf{X}_{k,-i}^*)- \mathcal{M}(X_{ij}^*=0|t_l,\mathbf{X}_{k,-i}^*)||_2^2$
	}
\end{algorithm}

otherwise.

The CPF resembles the squared gradient of a multivariate function. To approximate the gradient well, $\delta$ needs to be small relative to the range of $\mathbf{X}$. Large $J$, $n$ and $n_t$ can yield a more accurate estimation of the predictive ability of the features; yet, the choice of these parameters may be constrained by the computing resources available. 

\begin{figure*}[t!]
	%	\begin{subfigure}{\linewidth}
		\centering
		
		\subfloat{
			\raisebox{0.005\textwidth}{\small \hspace{3cm} Linear Relationship \hspace{5cm} Nonlinear Relationship} 
		}
		
		\subfloat{
			\raisebox{0.12\textwidth}{\rotatebox[origin=c]{90}{\small Continuous}}   %%% 0.75in is half of figure height 1.5in
		}
		\subfloat{\includegraphics[width=0.485\textwidth]{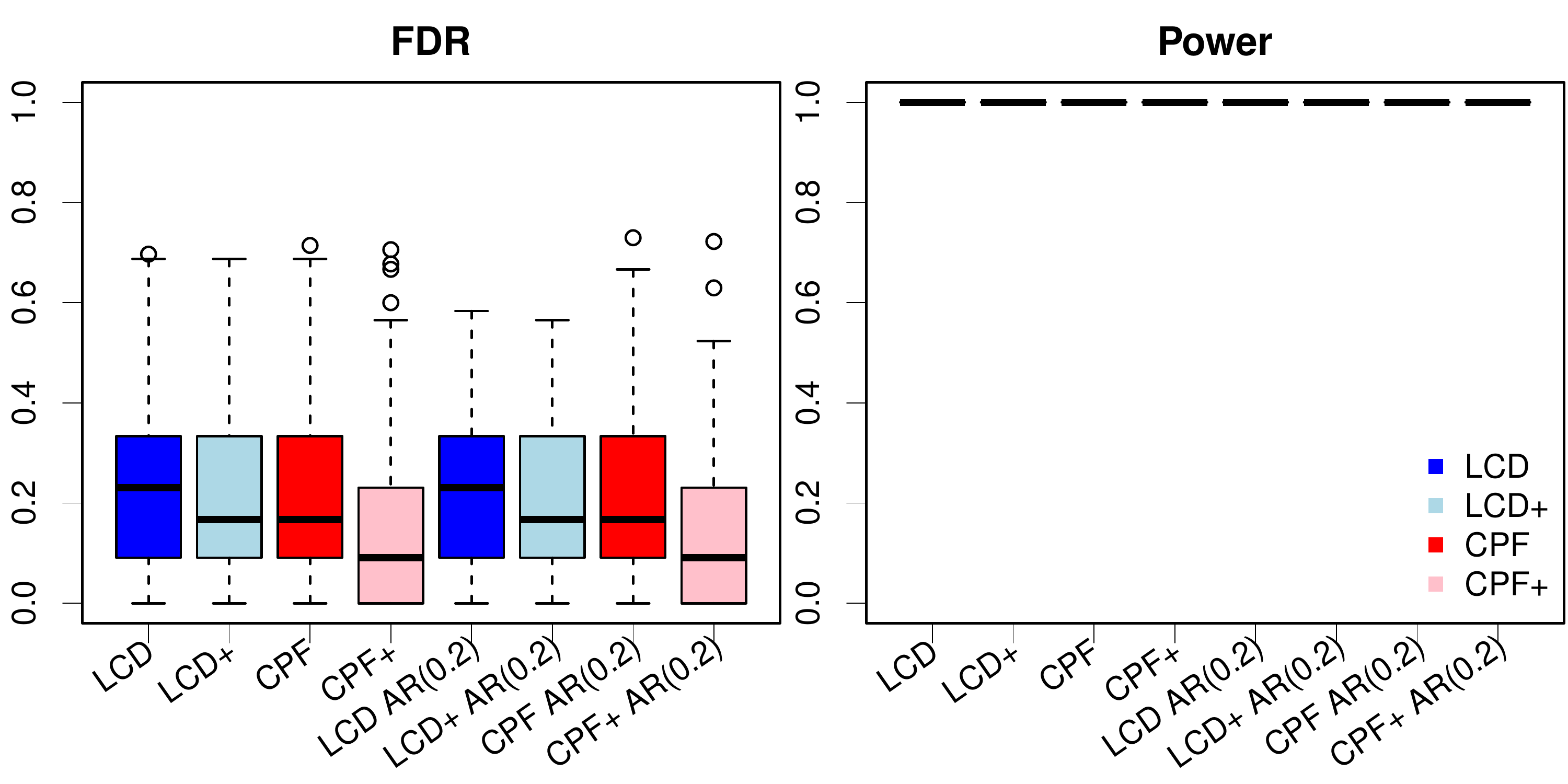}
			\includegraphics[width=0.485\textwidth]{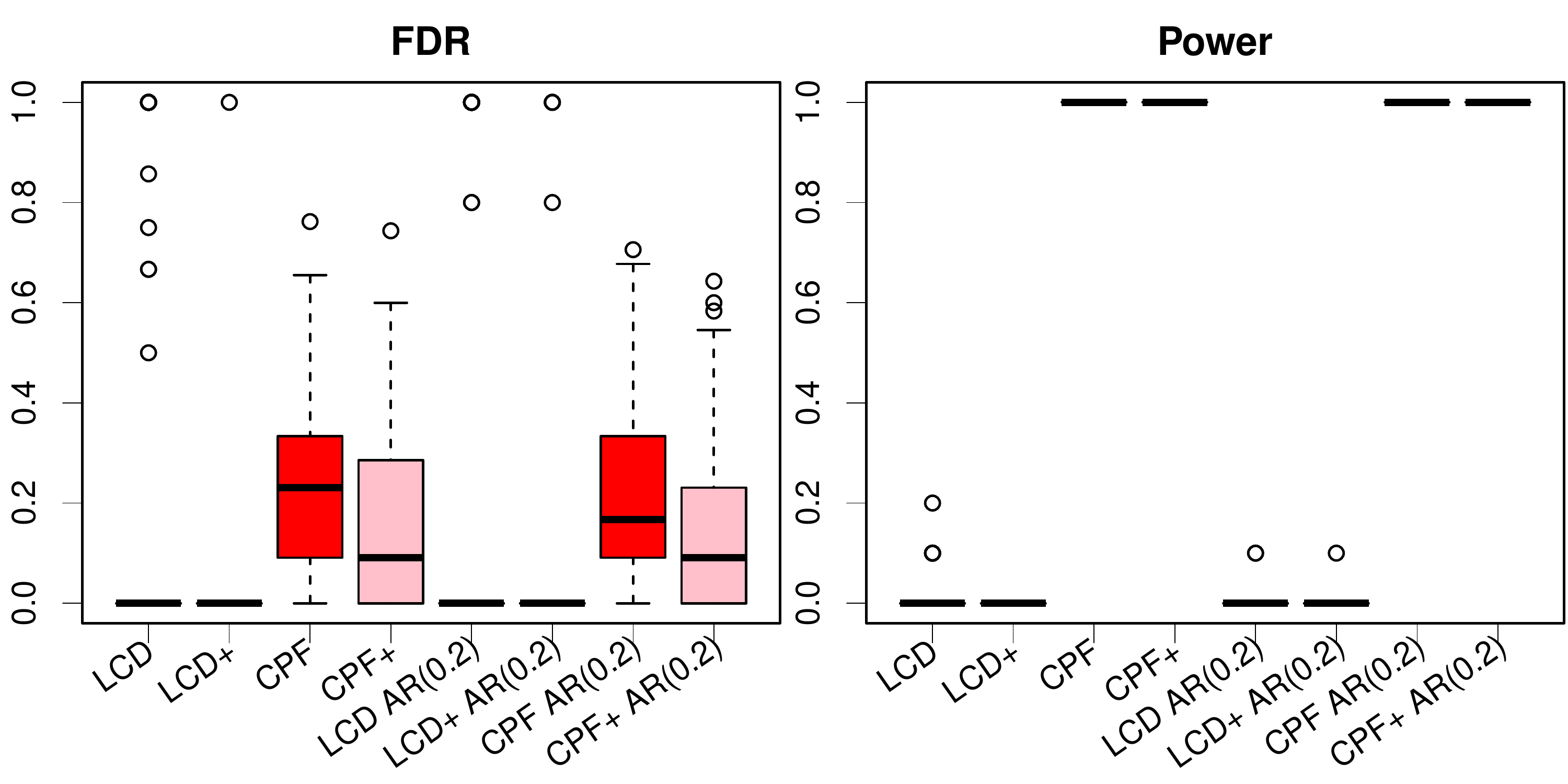}}
		
		\subfloat{
			\raisebox{0.125\textwidth}{\rotatebox[origin=c]{90}{\small Binary}}   %%% 0.75in is half of figure height 1.5in
		}   
		\subfloat{\includegraphics[width=0.485\textwidth]{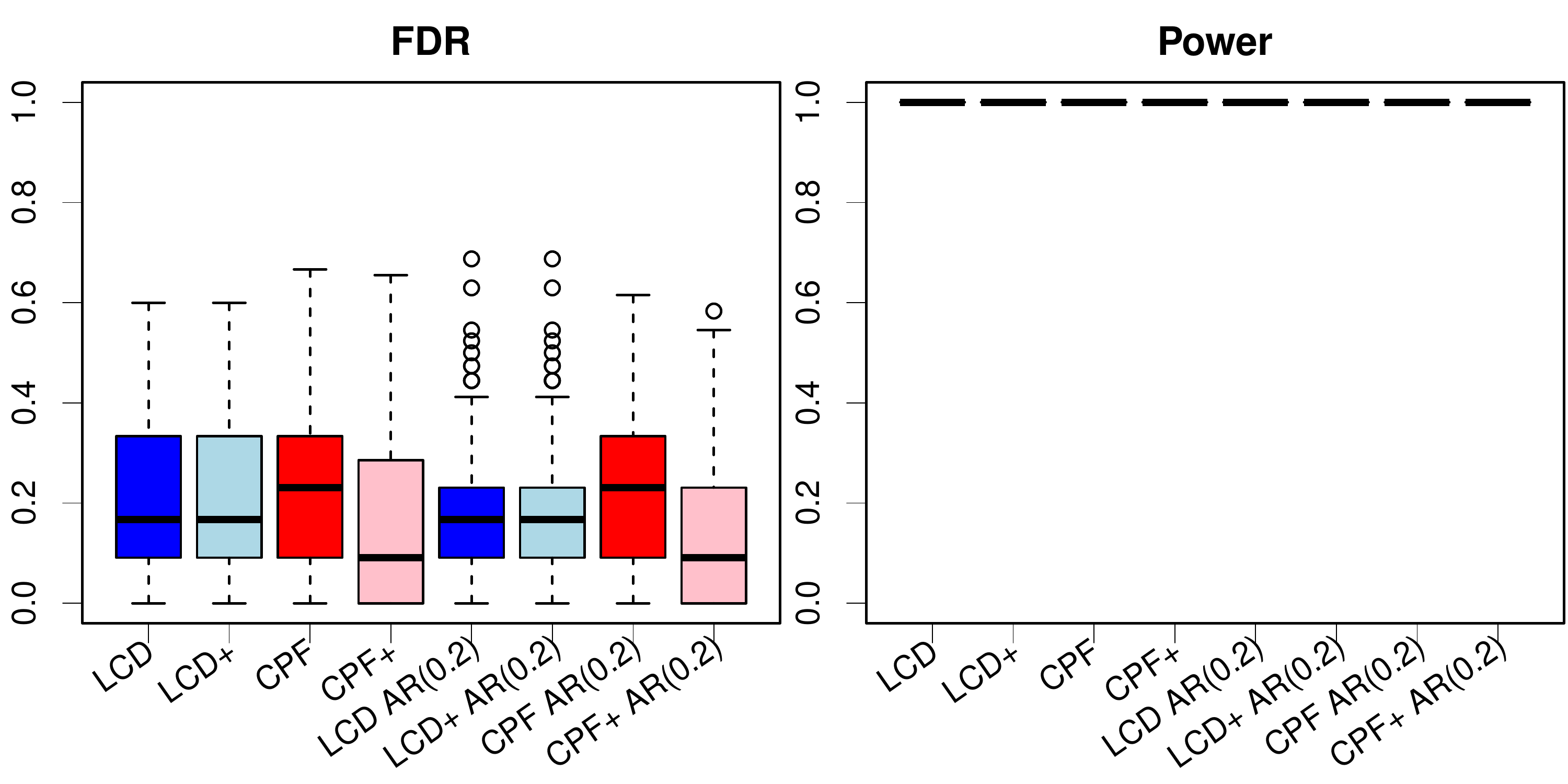}
			\includegraphics[width=0.485\textwidth]{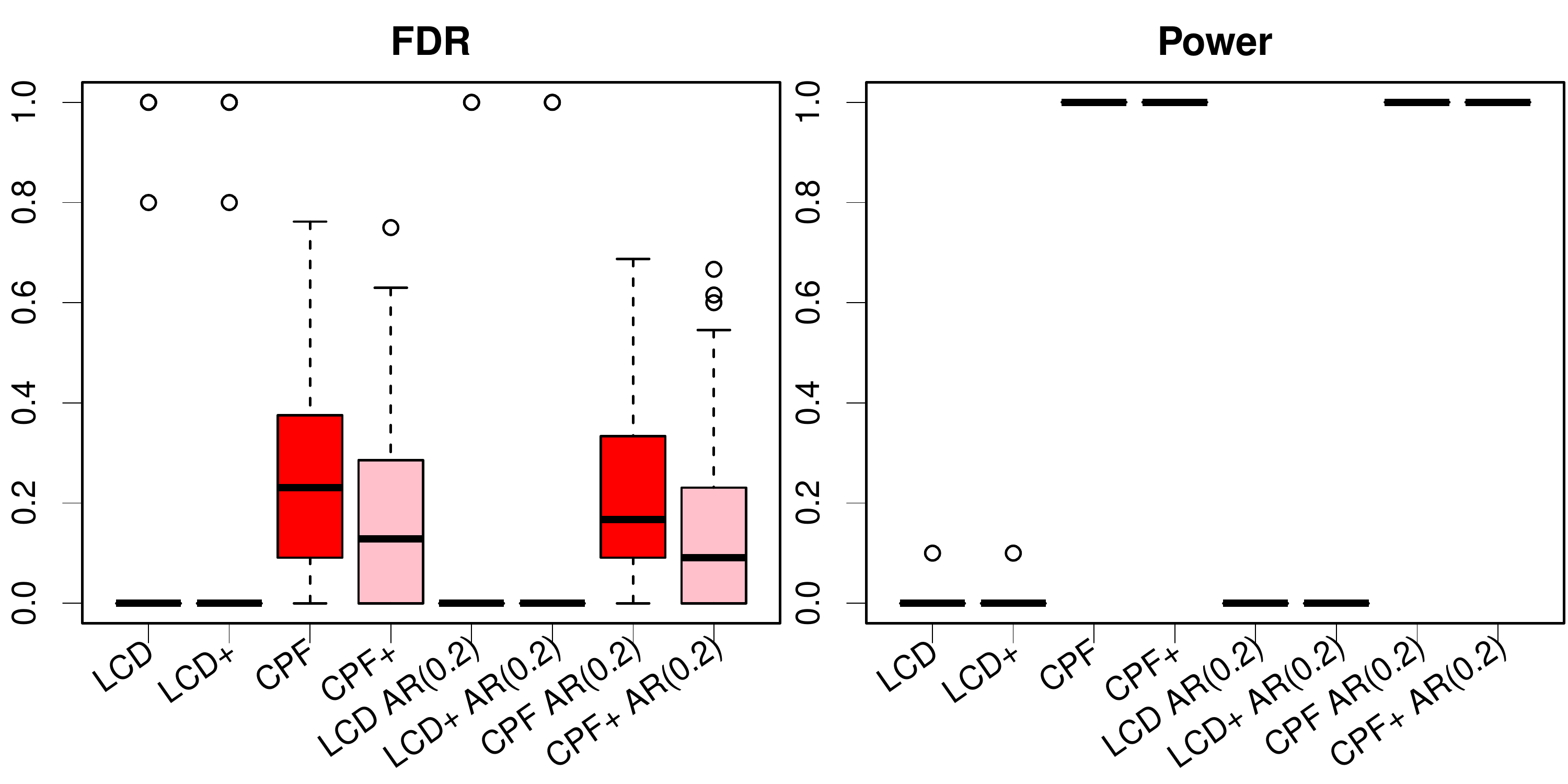}}
		
		\subfloat{
			\raisebox{0.005\textwidth}{\small \hspace{1.0cm} Log-linear Hazard Relationship \hspace{3.5cm} Non Log-linear Hazard Relationship} 
		}       
		
		\subfloat{
			\raisebox{0.125\textwidth}{\rotatebox[origin=c]{90}{\small Survival}}   %%% 0.75in is half of figure height 1.5in
		}   
		\subfloat{\includegraphics[width=0.485\textwidth]{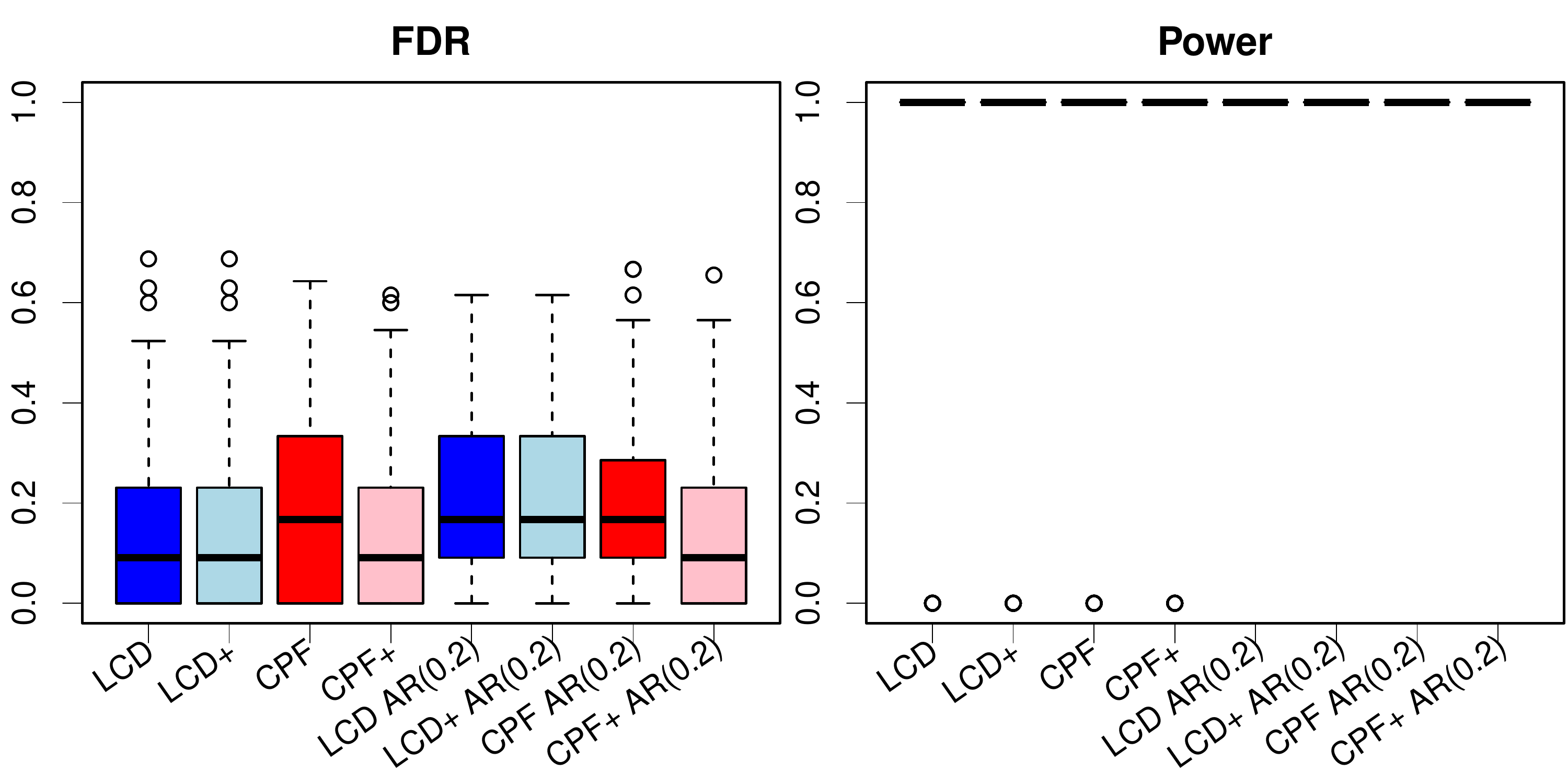}
			\includegraphics[width=0.485\textwidth]{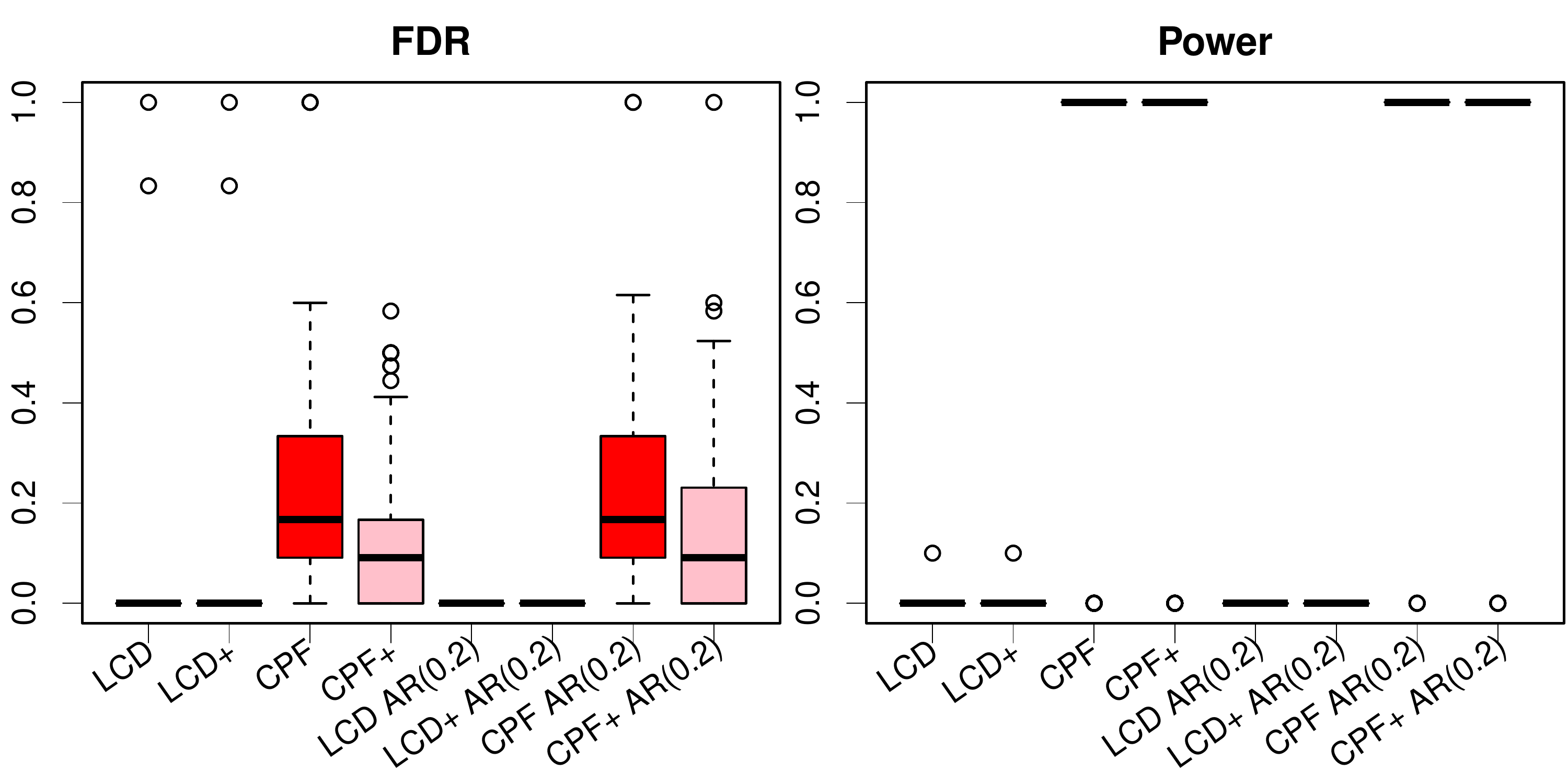}}
		
		\subfloat{
			\raisebox{0.125\textwidth}{\rotatebox[origin=c]{90}{\small Competing risks}}   %%% 0.75in is half of figure height 1.5in
		}   
		\subfloat{\includegraphics[width=0.485\textwidth]{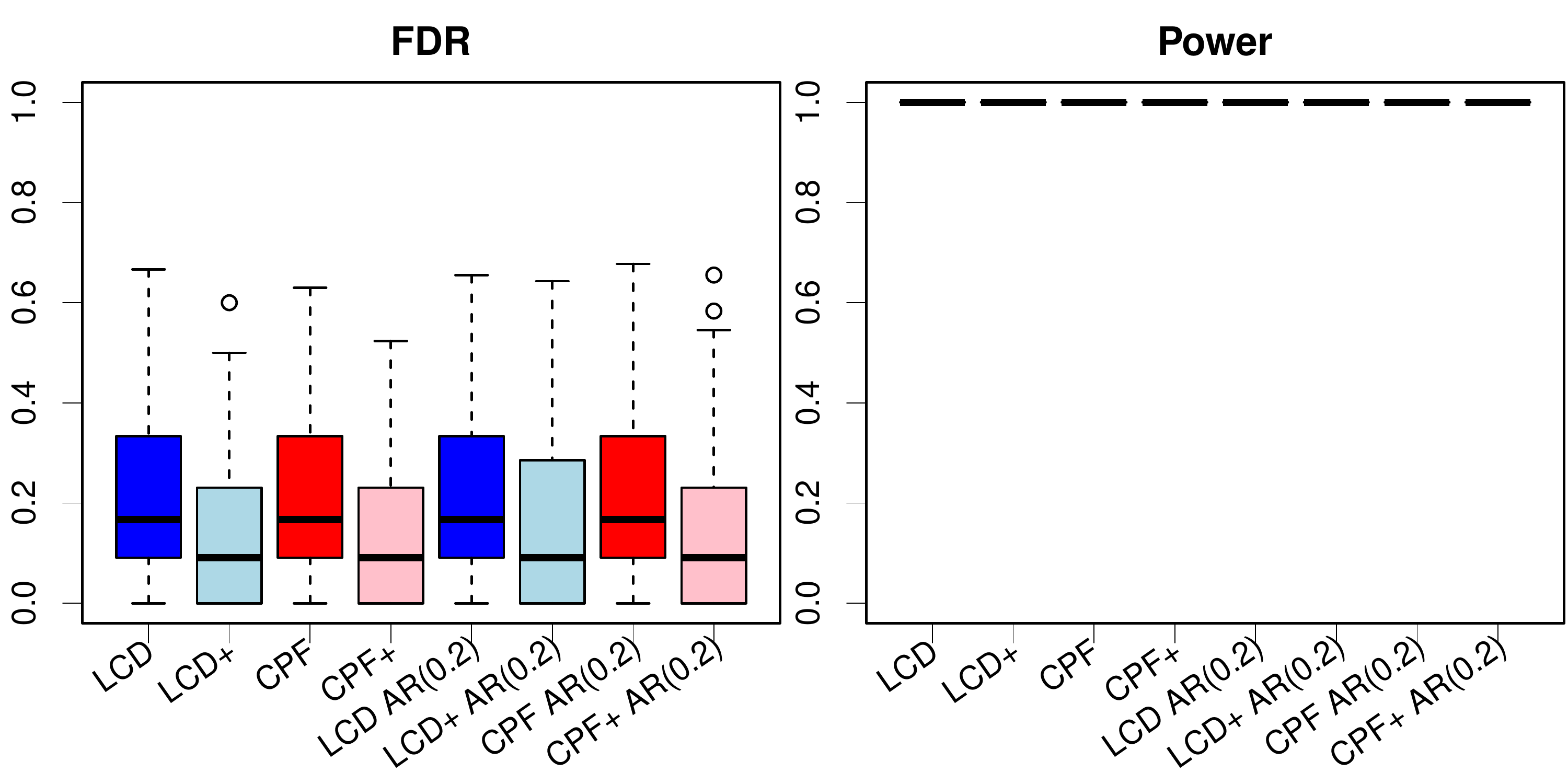}
			\includegraphics[width=0.485\textwidth]{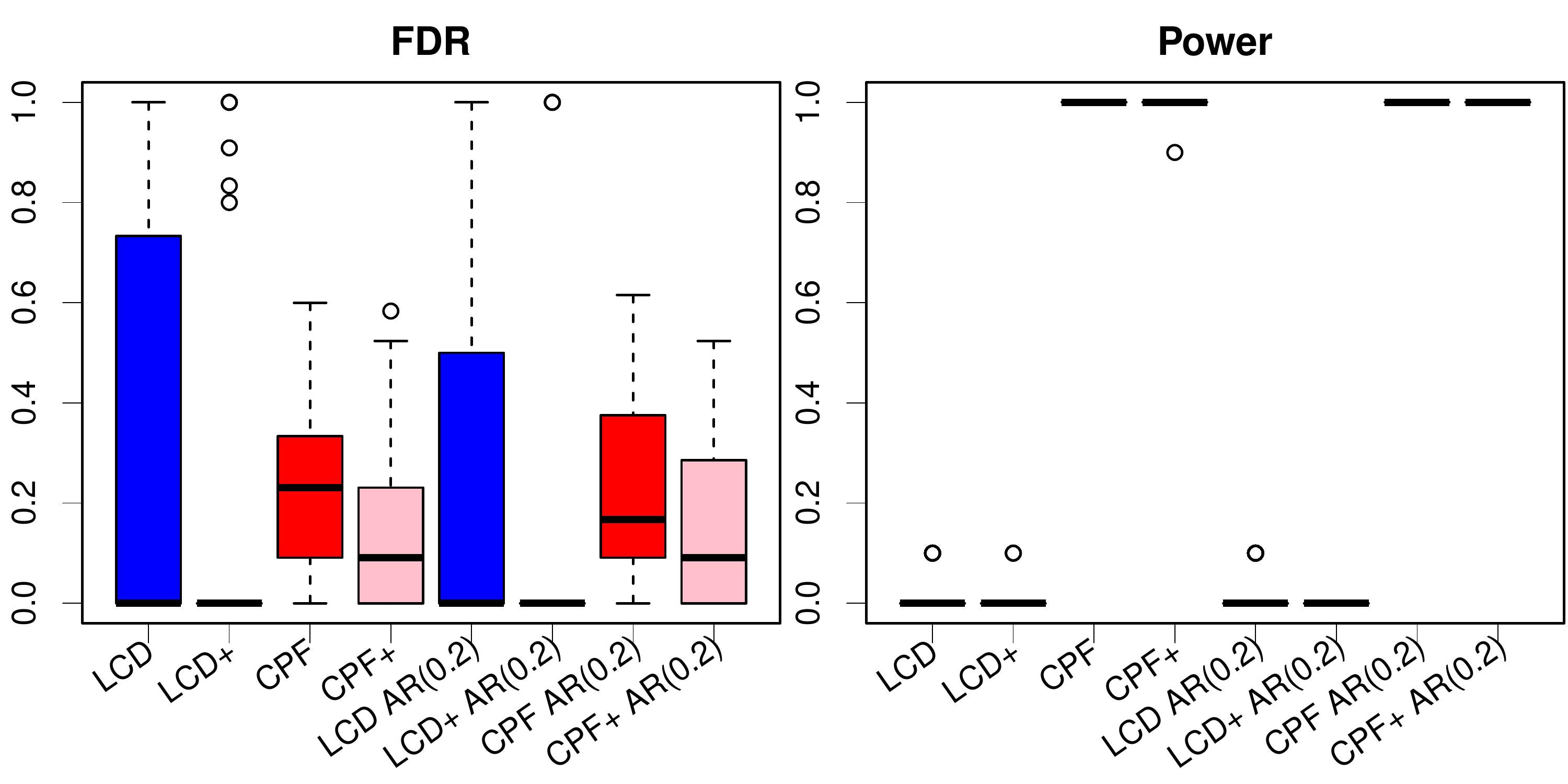}}
		\\
		\caption{Comparison of the FDR and power for CPF and LCD statistics. LCD and CPF denote standard knockoff filtering using the LCD and CPF statistics; LCD+ and CPF+ denote knockoff+ filtering for these techniques. 
  The left columns show the results when the relationship between covariates and the outcome is linear/log-linear, while the right two columns show the results when the relationship is not.}
		\label{simulation}
	\end{figure*}
\section{SIMULATION STUDY}
\label{sec:Simulation}
To illustrate the ability of the CPF statistics to detect both linear and nonlinear effects of predictors, we evaluated their performance in conjunction with DNNs for various outcome types, including continuous, categorical, survival, and competing risks data. In all simulated datasets, the number of original features was set at $250,$ while the sample size was $20000.$ Among the $250$ predictors, ten were genuinely predictive. During the training of DNNs, early stopping callback was implemented to mitigate overfitting, with the patience parameter set at 50 for continuous, categorical, and survival outcomes and at 200 for competing risks data. For continuous and categorical outcomes, a batch size of 20 was employed, and the validation proportion in each epoch was designated at 10\%. For survival and competing risks outcomes, a batch size of 50 was used, and the validation proportion in each epoch was set at 20\%. The parameters $J=5$, $n=100$, and $\delta=0.1$ were specified for the calculation of the CPF statistics from fitted DNNs.

The LCD statistics were selected as the benchmark for comparison, as it has demonstrated higher power than the LSM statistics in most instances \citep{CandFan18}. For both continuous and categorical outcomes, the LCD statistics were derived from lasso results; for survival outcomes, they were based on Cox lasso results, and for competing risks outcomes, they were obtained from the Fine-Gray regression model \citep{FineGray99}. 

The target false discovery rate was set to be $0.2.$ Original features were generated using either a standard normal distribution or an autoregressive model with a mean of $\mathbf{0}$ and a covariance matrix with $0.2^{|i-j|}$ entries, where $i$ and $j$ denote column and row indicators, respectively. The covariance matrix of the autoregressive model contains $1$ along the diagonal, with entry values diminishing to zero as they move further from this diagonal. The second-order approximate Gaussian knockoffs, provided by the R package ``knockoff" \citep{PattSesi20}, were employed as the knockoff generator. Each simulation configuration was repeated $200$ times, and the results were summarized using box plots. Subsequent subsections will present the simulation setup for each type of outcome.

\subsection{Continuous Outcome}
For continuous outcomes, the true distribution of the response variable was assumed to follow a linear model with predictors that are either linear or quadratic functions of X. In the linear setting, the predictors and the outcome are related as follows: $y_i=\boldsymbol{\beta}^T \mathbf{X}^*_{i} +100+\epsilon_i,$ $i=1,2,\dots,n,$ where $\mathbf{X}^*_i$ represents the vector that includes only true predictors, and $\boldsymbol{\beta}$ is a vector with all entries equal to 
3. In the nonlinear setting, the relationship between each predictor and the outcome is quadratic, given by: $y_i=\sum_{j=1}^{10}\mathbf{X}^{*2}_{ij},$ $i=1,2,\dots,n.$ 

In the linear setting, a three-layer DNN was employed, where the internal layer consisted of 8 nodes. For the nonlinear setting, a five-layer DNN was used, with the numbers of nodes in internal layers specified as (64, 32, 16). The activation function for both models was chosen to be the Rectified Linear Unit (ReLU), and for each layer, an L1 regularization term was imposed with a parameter value of 0.001. The optimizer was configured to be Adam, with a learning rate set at 0.01, and the loss function was defined as the mean squared error in the validation data.

\subsection{Categorical Outcome}
The data simulation for categorical outcome is similar to the simulation for continuous outcomes. In the linear relationship setting, the predictors and the expected value of the outcome follow a linear relationship on the logit-transformed scale according to
$y_i \sim \operatorname{binom}(\operatorname{logistic}(\boldsymbol{\beta}^T \mathbf{X}^*_{i})),$ $i=1,2,\dots,n.$ $\mathbf{X}^*_i$ is the vector including only true predictors and $\boldsymbol{\beta}$ is a vector of $1's.$ In the nonlinear setting, we first computed the sum of squares of the true predictors $Z_i=\sum_{j=1}^{10}\mathbf{X}^{*2}_{ij},$ then standardized $Z_i$ and generated the outcome  
$y_i \sim \operatorname{binom}(Z_i).$ 

To decrease the training time, we started with the model trained for continuous outcomes and simply changed the last layer to a sigmoid activation function. The loss function was chosen to be binary-crossentropy in the validation data. The learning optimizer was chosen to be Adam, and the learning rate was 0.01.

\subsection{Survival Outcome}
We also applied our method on data with survival outcomes. We used DeepSurv model \citep{KatzShah18}, which adopts the framework of the Cox proportional hazards model, for computing the predicted survival based CPF. In the DeepSurv model, the hazard ratio of an individual is parameterized by the weights of the network, and the optimization goal is to maximize the log partial likelihood with some regularization.
Again, we use $X_j^*,\quad j=1,\dots, 10,$ to denote the $j$th true predictor and specify that the hazard for each individual follows $h_i(t|\mathbf{X}) = \exp(\sum_{j=1}^{10} 0.5X_{ij}^*),\quad i=1,\dots,n.$ This satisfies the underlying assumption for the Cox model as the covariate imposes a time-invariant multiplicative effect on the hazard. We refer to the scenario in which the covariate affects the hazard through a log-linear relationship as the log-linear setting. The censoring times were generated independently from an $\exp(0.05)$ distribution and this resulted in a $5\%$ censoring rate. In simulation scenarios where the covariate effects are not log-linear, the hazard for the $i$th individual was $h_i(t|\mathbf{X}) = \exp(\sum_{j=1}^{10} 0.5X_{ij}^{*2}),$ and the censoring times were generated independently from an $\exp(0.1)$ distribution. %Here, the covariate effects are not log-linear and is called the non log-linear setting.
	
For the scenarios under the log-linear setting, the numbers of nodes were set to be $(64, 32, 16, 8, 1),$ for the non log-linear scenarios, the numbers of nodes were $(128, 64, 32, 16, 1).$ Batch normalization and a dropout rate $0.5$ were employed. We used Adam optimizer with learning rate 0.0001 and the weight decay was set to be 0.01.
	
\subsection{Competing Risks Outcome}
To further assess the capabilities of the CPF statistics, we explored the application with competing risks data. In competing risks data, there are several types of potential events and the occurrence of one type of outcome prevents us from observing other types of events. Suppose that type 1 events are the main interest. \cite{FineGray99} mimicked the Cox proportional hazards model by transforming the CIF $F_1(t)$ for type 1 into a subdistribution hazard $\eta_1(t),$
\begin{equation}
	\eta_1(t) = -\frac{\partial}{\partial t} \log(1-F_1(t)),
	\label{cmprsk1}
	\end{equation}
and assumed the covariates have multiplicative, time-constant effects on the subdistribution hazard through the form
\begin{equation}
	\eta_1(t|\mathbf{x}) = \eta_{01}(t) \exp\left(f(\mathbf{X})\right).
	\label{cmprsk2}
\end{equation}

Recently, \cite{LeeZame18} leveraged DNN for modeling competing risks data and proposed the DeepHit model, which was employed in computing the predicted CIF used in CPF statistics in our simulation.
	
Our simulations assumed that 2 event types exist and that type 1 is the primary interest. In both log-linear and non log-linear hazards scenarios, we assumed the baseline probability of experiencing event type 1 was $q=0.5.$ The influence of covariates on outcomes was encapsulated by the variable $Z = f(\mathbf{X})$. Specifically, in the log-linear scenario, $Z_i=\sum_{j=1}^{10} X_{ij}^*,\quad i=1,\dots,n.$ For the non log-linear scenario, $Z_i=\sum_{j=1}^{10} X_{ij}^{*2}-10.$  For both scenarios, $Z_i$ is connected to the subdistribution hazard function through equation \ref{cmprsk2}. By virtue of (\ref{cmprsk1})-(\ref{cmprsk2}), the relationship between the baseline and individual CIFs is
\begin{equation}
	F_1(t|\mathbf{X}_i) = 1 - (1 - F_{01}(t))^{\exp(Z_i)}.
	\label{cmpgen}
\end{equation}
	
Generating event times involved a two-step process: first, we generated probabilities of experiencing a type 1 event for each subject by randomly sampling from a binomial distribution with a probability of $F_1(\infty|\mathbf{X}_i) = 1 - (1 - q)^{\exp(Z_i)};$ subsequently, for those subjects experiencing type 1 events, we modeled the baseline CIF for event type 1 as $F_{01}(t) = q(1 - \exp(-t))$ and generated their event time from the cumulative distribution function $F_1(t|\mathbf{X}_i)/F_1(\infty|\mathbf{X}_i)$, where the numerator follows (\ref{cmpgen}). For type 2 events, we generated event times from an exponential distribution with a rate parameter of 1. Censoring times were independently generated from an $\exp(1)$ distribution.
	
Within the architecture of the DeepHit, for both the log-linear and non lig-linear scenarios, the number of internal nodes was configured as $(128, 64, 32, 16).$ To enhance training performance, we applied batch normalization and utilized a dropout rate of 0.5. In fitting the DeepHit model, the number of cutpoints for evaluating the likelihood was set to be 20. 
	%We prioritized likelihood maximization by setting loss ratio $\alpha=1.0.$ 
To mitigate overfitting, an L2 penalty with a parameter of 0.1 was incorporated.

\subsection{Summary of Simulation Results}
As shown in Figure \ref{simulation}, the simulation results are consistent across different outcome types.
When the linear assumption or the log-linear assumption is satisfied, the performance of the CPF is similar to the performance of the linear model-based LCD. However, when this assumption does not hold, for example, the relationship between the predictors and outcome is quadratic, the LCD fails to capture the true relationship and exhibits significant power loss, while the CPF controls the FDR while offering high power to detect signals.
	
\section{REAL DATA APPLICATION}
	
\label{sec:Application}
	
	\begin{table*}[t!]\centering
		
	\caption{Genes identified by the CPF method. Genes that corroborate with the selection from LCD statistics results are shown in bold.}
	\begin{tabular}{ll}
		\textbf{GENE}&\textbf{RELEVANT LITERATURE}\\\hline
		\textbf{GAPDH}&\cite{PuzoSava13,LiuTang17}\\
		\textbf{RPLP0P6} &\cite{XieLin23,WangQian18}\\ %\citep{KouhAzim19}(bladder cancer) 
		RNASE1&\cite{TravNabh20,LiYang19,MiroVlas19}\\
		SFTPC&\cite{LiTodd07,WangPhil09,LiMeng19}\\
		%PPIAP22&\citep{ChenLuo16}\\
		%MT-TN 
		%MT-ND4L
		IGKV4-1&\cite{WangWang20,QinXu20,XiaWu22,PedeJens22}\\
		IGHJ3&\cite{IsaeShar19,FengXiao09,JianZhou22} \\
	\end{tabular}
		
	\label{literatureGene}
\end{table*}

\subsection{Identifying predictors of real estate sale price}
We analyzed a dataset \citep{RafiAdel18} consisting of 108 covariates, aiming to identify predictors of the sale prices of single-family residential apartments in Tehran, Iran. The dataset encompasses various factors, including temporal economic covariates
%aspects 
as well as diverse project-related physical and financial attributes. 
%These attributes encompass the project's locality indicated by zip codes, the building's total floor area, lot area, initial estimated construction cost based on project commencement prices, corresponding estimated costs in a selected base year, construction duration, unit price at project initiation, and actual construction expenses. 
Furthermore, the dataset includes economic indicators and indices across five periods prior to the start of construction, such as 
stock market index and city population.
%	the count of building permits issued, building services index relative to a chosen base year, wholesale price index for building materials in the base year, total floor area for permits granted by the municipality, private sector investment in new constructions, base year land price index, bank loan quantities (amount and number) and interest rates, average private sector construction costs at project completion and initiation, official and unofficial exchange rates against the dollar, consumer price index both overall and specific to housing, energy, and utilities, stock market index, city population, and gold price per ounce.
	
We divided the 372 observations into training ($70\%$) and testing ($30\%$) sets.
%, assigning 30\% to the testing subset. 
We first fit the lasso model with cross-validation, yielding a mean squared error of 27767.39 on the testing dataset. Then, we employed the LCD knockoff statistics with a target false discovery rate of 0.2. The knockoff+ method led to no feature selection, while the standard knockoff procedure resulted in the selection of a single feature: the unit price at project initiation.
	
To compute CPF knockoff statistics, we started by constructing a DNN with four layers. The internal layers contained 256 and 128 nodes respectively, utilizing the leakyReLU activation function with an alpha parameter of 0.3. The final layer adopted the ReLU activation function. Across all layers, batch normalization was implemented. We opted for the Adam optimizer with a learning rate of 0.01 and chose batch size to be 50. To prevent overfitting, early stopping with a patience parameter of 200 was employed. The DNN can achieve a mean squared error of 20787.06 on the test dataset, which outperforms lasso in model fitting. When adopting the more conservative knockoff+ approach, similar to the result from LCD statistics, no features were selected. Meanwhile, with the standard knockoff method, three features were chosen: equivalent preliminary estimated construction cost based on prices at the project's outset, unit price at project initiation, and actual construction expenses.
	
\subsection{Determinants of lung adenocarcinoma stage from TCGA dataset}
We applied CPF statistics on lung adenocarcinoma gene expression data from The Cancer Genome Atlas (TCGA). Our primary focus was on determining genes that express differently for stage I and stages II/III/IV cancer. In this dataset, there were 330 patients with stage I cancer and 271 patients with stages II/III/IV cancer. From the original 31,294 features, considering the constrained sample size, we selected 100 features with the highest variance. The target FDR here is 0.2. Before fitting the model, the predictors were standardized. We allocated $70\%$ of the data for training and utilized the remainder for testing the model. Given the limited sample size, we employed a three layer neural network, with an internal layer with 100 nodes. The activation functions were chosen to be ReLU and sigmoid. Training DNN involved using dropout (dropout rate = 0.2) and L1 regularization with penalty parameter of 0.1 for kernel and 0.01 for bias in the last layer. The batch size was chosen to be $100.$ The Adam optimizer with a learning rate of $0.001$ was used, and early stopping call with patience parameter 200 was employed to prevent overfitting. 
	
Using cross-validation, the lasso model yielded a c-index of 0.60088 for the test data, while the DNN achieved a c-index of 0.6774, demonstrating a superior model fitting than the lasso model. The knockoff+ method did not select any features for either LCD or CPF knockoff statistics. However, with the regular knockoff method, 6 features were chosen using the CPF approach, and LCD selection corroborated two of these (LCD with lasso chose 3 features in total). Table \ref{literatureGene} lists all the genes selected by CPF and recent biomedical publications that have identified the correlations between the same gene and lung cancer.

In both examples, the CPF statistics selected more features than LCD statistics. This behavior can be explained by its capacity in capturing nonlinear relationship between covariates and outcomes. The real data application is consistent with simulation study outcomes, demonstrating that CPF statistics have better power than a widely used existing method.

\section{DISCUSSION}
\label{sec:Discussion}
Applying the CPF statistics successfully relies on pairing them with a well-selected and trained machine learning model. This paper is not aimed at providing a one-size-fits-all solution for variable selection and the flexibility of the CPF does not eliminate the need for tuning complicated machine learning models, in particular DNNs. We encourage users to first check the model fitting metrics of the trained model, including mean squared error, prediction accuracy, and concordance index and comparing them to the performance of traditional methods, such as linear model, lasso, and the Cox model. After verifying that the machine learning model attains better model fitting, then CPF can be used with it to identify key features that are predictive of the outcomes.  
	
Complex machine learning models such as DNNs have long been perceived as “black boxes,” such that the relationships between features and response variables are hard to discern from the fitted model and which variables are truly associated with the outcome is unclear. The CPF knockoff statistic offers improved model interpretability by capturing the nonlinear relationships between the covariates and outcomes and enabling these correlations to be identified through knockoff filtering.  When applied in biomedical studies, CPF statistics offer an improved resolution of prognostic features for health outcomes. This improvement in feature identification will promote the understanding of the determinants of patient health and, subsequently, can lead to advances in the prevention, treatment, and management of disease, thereby improving public health. Also, The proposed technique will be helpful more broadly in other scientific contexts such as information technology (IT), engineering, sociology, economics, psychology, and environmental science, where a large set of candidate variables are being evaluated for association with an outcome of interest.

\section*{ACKNOWLEDGMENT}

This project is partly supported by NSF DMS 2310955. We would like to thank Dr. Hao Feng of Case Western Reserve University for his help in TCGA data application.

\bibliographystyle{apa}

\end{document}